\documentclass[11pt]{article}
\usepackage[a4paper,margin=1in]{geometry}
\usepackage[T1]{fontenc}
\usepackage{mathptmx}
\usepackage{amsmath,amssymb,booktabs,graphicx,setspace,float,tabularx,array,etoolbox,adjustbox}
\usepackage[final]{microtype}
\usepackage[labelfont=bf,labelsep=period,font=small,skip=6pt,justification=centering]{caption}
\usepackage{titlesec}
\titleformat{\section}{\large\bfseries}{\thesection}{0.8em}{}
\titleformat{\subsection}{\normalsize\bfseries}{\thesubsection}{0.8em}{}
\titleformat{\paragraph}[runin]{\normalsize\itshape}{}{}{}[.]
\titlespacing*{\section}{0pt}{20pt plus 4pt minus 2pt}{8pt plus 2pt}
\titlespacing*{\subsection}{0pt}{14pt plus 3pt minus 2pt}{6pt plus 2pt}
\titlespacing*{\paragraph}{0pt}{6pt plus 2pt}{0.6em}
\usepackage[authoryear,round]{natbib}
\usepackage{xcolor}
\definecolor{linkblue}{RGB}{0,51,102}
\usepackage[colorlinks=true,citecolor=linkblue,urlcolor=linkblue,linkcolor=linkblue,pdfauthor={},pdfcreator={}]{hyperref}
\usepackage[section]{placeins}
\graphicspath{{figures/}}
\AtBeginEnvironment{table}{\small}
\newcommand{\CSCI}{\mathrm{CSCI}}
\newcolumntype{L}[1]{>{\raggedright\arraybackslash}p{#1}}
\newcommand{\tabnotes}[1]{\par\vspace{5pt}\begin{minipage}{\linewidth}\footnotesize\textit{Notes:} #1\end{minipage}}
\newcommand{\fittable}[1]{\begin{adjustbox}{max width=\textwidth}#1\end{adjustbox}}

\newcommand{\paperabstract}[3]{%
  \begin{center}\begin{minipage}{0.88\textwidth}\small
  \noindent\textbf{Abstract}\par\smallskip\noindent #1\par\medskip
  \noindent\textbf{Keywords:} #2\par\smallskip
  \noindent\textbf{JEL classification:} #3
  \end{minipage}\end{center}}

\hypersetup{pdftitle={When Values-Based Screens Disagree: Continuous Shariah Compliance and Classification Dynamics},
  pdfauthor={Abdulrahman Qadi, Akash Sharma, Francesca Medda}}

\begin{document}
\thispagestyle{empty}
\begin{center}
{\LARGE\bfseries When Values-Based Screens Disagree:\\[4pt] Continuous Shariah Compliance and Classification Dynamics\par}
\vspace{16pt}
{\large Abdulrahman Qadi\textsuperscript{*} \qquad Akash Sharma \qquad Francesca Medda\par}
\vspace{6pt}
{\normalsize Institute of Finance and Technology, University College London\par}
\vspace{10pt}
\end{center}
\vspace{10pt}
\paperabstract{Does the design of a values-based screening rule generate eligibility changes when reported accounting inputs remain unchanged but market valuations move? We apply seven Shariah financial-screening rulebooks to 1.34 million U.S. security-months from 1999 to 2024 on a common, threshold-anchored scale. Applied retrospectively, the full emulations disagree on 35.9\% of complete permissible-sector observations (38.8\% from 2015, when all were in force), and the financial-ratio screens alone on 47.3\%. Under AAOIFI's spot market-capitalisation denominator, market-capitalisation movement alone is sufficient for 65.6\% of breaches, and half of denominator-driven breaches reverse within three months; numerator movement dominates under asset-based rules. Holding the threshold and sample fixed, a 24-month average denominator cuts debt-screen transitions from 2.51\% to 1.03\% a month, whereas raising the threshold barely matters. Among breaches of similar severity the reversal gap narrows but persists. Portfolio and fund-holdings evidence qualifies these findings: benefits are modest under a binding single-rule mandate.
}{sustainable finance; values-based investing; Shariah screening; screening-rule design; classification stability; portfolio turnover}{G11, G23, C43, Z12}
\begingroup\renewcommand{\thefootnote}{}\footnotetext{\textsuperscript{*}Corresponding author: \href{mailto:abdulrahman.qadi.24@ucl.ac.uk}{abdulrahman.qadi.24@ucl.ac.uk}. Akash Sharma: \href{mailto:akash.sharma@ucl.ac.uk}{akash.sharma@ucl.ac.uk}. Francesca Medda: \href{mailto:f.medda@ucl.ac.uk}{f.medda@ucl.ac.uk}. We are grateful to participants at the UCL Institute of Finance \& Technology research seminars for helpful comments and suggestions. We thank WRDS for providing access to CRSP and Compustat data under institutional licence. Any remaining errors are our own. Earlier versions of this paper circulated under the title ``From Binary Screens to Continuous Compliance: A Shariah Screening Measure for Portfolio Design.''}\endgroup
\newpage
\setcounter{page}{1}

\section{Introduction}

Values-based investors increasingly delegate eligibility decisions to codified screens. Environmental, social and religious mandates alike turn continuous company data into a yes-or-no permission, and the permission then drives index membership, fund holdings and trading. A screen's design therefore matters beyond its intended values. If a rule's arithmetic makes eligibility switch on and off while the accounting data it measures are unchanged, investors bear turnover and uncertainty that come from market prices rather than from the firm's reported activities. Some standard setters may intend that price sensitivity; this paper measures its consequences and does not judge the objective. This paper asks a direct empirical question: does the design of a values-based screening rule generate eligibility changes when reported accounting inputs remain unchanged but market valuations move?

Shariah equity screening is an unusually transparent setting in which to answer it. Major rulebooks publish numerical financial screens but differ in thresholds, components, denominators and averaging windows \citep{KhatkhatayNisar2007,RizaldyAhmed2019}. Some divide debt by spot or trailing market capitalisation; others by total assets. These choices alter eligible universes and portfolio characteristics \citep{MarzbanAsutay2012,Raza2021}. Because several codified rules map the same balance sheet and market prices into different permissions, the setting allows the effect of rule design to be separated from changes in reported accounting inputs.

We emulate seven financial-screen specifications from five standard setters or index-provider families on a monthly panel of U.S. securities, with dated rule changes and conservative accounting timing. For every compliance change we ask which input moved: the numerator (debt, cash, receivables or income), the denominator, or both. We then hold the threshold and sample fixed while changing only the denominator. To compare rulebooks on one scale, we map each ratio relative to the threshold of the rule that uses it into a Continuous Shariah Compliance Index (CSCI). CSCI is the measurement framework that organises the evidence, not the paper's main claim.

Three findings stand out. First, rulebooks disagree often. Applied retrospectively to common inputs, the seven specifications disagree on 35.9\% of complete permissible-sector observations, and on 38.8\% from September 2015, when all seven were in force. On the financial-ratio screens alone, without the approximate income test, they disagree on 47.3\%. Second, under AAOIFI's spot market-capitalisation denominator, most compliance changes are price events. Market-capitalisation movement alone is sufficient for 65.6\% of compliant-to-non-compliant transitions and 72.9\% of recoveries, and 39.2\% of breaches occur with every accounting input unchanged. These shares are similar when the income test is excluded. Numerator movement dominates under asset-based rules. These price-driven breaches are fragile: 49.7\% reverse within three months, against 11.1\% of breaches caused by the numerator. Part of the gap reflects that price-driven breaches are shallower, but among breaches of similar severity it remains about 15 percentage points. Third, the denominator design, not the threshold, drives instability. Holding the 30\% threshold and the sample fixed, moving from spot market capitalisation to a 24-month average cuts monthly debt-screen transitions from 2.51\% to 1.03\%, whereas raising the spot threshold from 30\% to 33\% leaves the transition frequency almost unchanged.

The remaining evidence qualifies these findings. CSCI orders next-month instability within binary states, but once the governing rule's own boundary distance is known, the multi-rulebook score adds little consistent information. Continuous tilts that may hold securities failing AAOIFI trade far less than a binary AAOIFI portfolio, but inside a binding AAOIFI mandate forced exits dominate turnover and CSCI lowers it by only 2.8 percentage points a year. In the holdings of two Shariah index ETFs and an active Shariah manager, emulated breaches of each fund's own rule coincide with more sales, but with only a small share of all sales.

The contribution is the joint measurement, decomposition and controlled comparison of screening-rule architectures on one panel. Prior work documents disagreement across Shariah standards, warns that market-capitalisation denominators make status sensitive to prices, and proposes averaging \citep{AyedhEtAl2019,ParlakEtAl2024}; recent work by \citet{Rai2026} quantifies the price channel for AAOIFI's leverage screen. We do not claim to have discovered price-driven screening changes. Table~\ref{tab:adds} sets out what this paper adds relative to each strand.

The paper also has clear limits. It does not show that smoothing is normatively preferable under Shariah governance; a longer average changes the rule and delays recognition of sustained deterioration. It does not show that CSCI predicts breaches better than the target rule's own margin. And it does not show that screening changes cause fund sales: quarterly holdings cannot establish trading sequence. The emulations approximate published rules and do not reproduce providers' databases or Shariah-board judgements. Finally, unchanged accounting records are not unchanged economic fundamentals: a price fall can reflect news about future profitability or distress before that news reaches financial statements. Our claims concern the inputs a rule measures, not firm value.

Section~\ref{sec:setting} describes the institutional setting and related work. Section~\ref{sec:data} describes the data, the screen reconstruction and CSCI. Sections~\ref{sec:disagreement}--\ref{sec:counterfactuals} report disagreement, the drivers of eligibility changes, and the architecture counterfactuals. Section~\ref{sec:portfolios} turns to portfolios and fund holdings, and Section~\ref{sec:conclusion} concludes with implications for values-based screening.

\section{Institutional setting and related work}\label{sec:setting}

\subsection{Screening rules}\label{sec:rules}

Shariah screening typically has two stages. Business-activity screens exclude prohibited core activities. Financial screens then limit interest-bearing debt, interest-bearing assets, receivables or liquidity, and non-permissible income. AAOIFI uses 30\% market-capitalisation limits for debt and cash-type components. DJIM and S\&P use trailing average market capitalisation, historically with additional cash and receivables screens. FTSE, MSCI's main series and the Securities Commission (SC) Malaysia use asset denominators; MSCI also maintains a market-capitalisation M-series. Our rules follow archived documents \citep{AAOIFI2024,DJIM2024,SP2024,MSCI2024,FTSE2022,SCMalaysia2013}. The rules change within the sample. DJIM and S\&P dropped their cash and receivables screens with the September 2023 rebalance \citep{SPDJIDJIM2023,SPDJIShariah2023}, and MSCI replaced a 70\% limit on receivables to assets with a 33.33\% limit on receivables plus cash between its September 2008 and August 2010 methodologies \citep{MSCI2008,MSCI2010}. Revisions continue after the sample: MSCI announced in late 2024 that it would raise its receivables-and-cash limit to 70\% \citep{MSCI2024consult}, and in September 2026 DJIM and S\&P lowered their debt limit to 30\% and reinstated a 30\% cash screen \citep{SPDJIDJIM2026,SPDJIShariah2026}.

The rulebooks also began at different dates: DJIM in February 1999, the AAOIFI standard on financial papers in May 2004, S\&P's Shariah indices in December 2006, MSCI's Islamic indices in July 2007, FTSE's Shariah series in January 2008, the SC's financial-ratio benchmarks in November 2013, and MSCI's market-capitalisation M-series in September 2015. Before those dates we apply each rulebook's formula retrospectively. The primary analysis is therefore a comparison of rule architectures under common inputs, not a reconstruction of the rules investors faced at each date; Section~\ref{sec:disagreement} reports the historically applicable alternative. The emulations are researcher reconstructions, not official constituent histories: public databases do not reveal provider-specific revenue classifications, case-by-case rulings or governance overrides, and some providers use daily prices with a reference-date share count where we observe month-end values \citep{SP2024}. The U.S. numerator labelled debt is predominantly conventional interest-bearing finance, so conclusions need not transfer to markets where firms can raise sukuk \citep{SukorHalim2022}.

\subsection{Related work}\label{sec:related}

\begin{table}[!htbp]
\centering
\caption{What this paper adds}\label{tab:adds}
\small
\begin{tabularx}{\textwidth}{L{2.6cm}XX}
\toprule
Dimension & Existing evidence & This paper \\
\midrule
Disagreement across standards & Different standards produce different eligible universes \citep{KhatkhatayNisar2007,DerigsMarzban2008,RizaldyAhmed2019}. & Monthly disagreement among seven dated rulebooks on one panel, separating retrospective and historically applicable comparisons (Section~\ref{sec:disagreement}). \\
Price-driven changes & Market-capitalisation denominators make status price-sensitive \citep{AyedhEtAl2019,ParlakEtAl2024}; \citet{Rai2026} attributes most AAOIFI leverage-screen flips in an S\&P 500 quarterly panel to price. & Decomposition of every screen component under seven rulebooks into price, shares, accounting and mixed causes, with a strict subset in which every accounting input is unchanged (Section~\ref{sec:drivers}). \\
Denominator smoothing & Averaging is proposed to stabilise status \citep{AyedhEtAl2019,ParlakEtAl2024}; existing comparisons change the threshold together with the denominator \citep{Rai2026}. & Threshold and sample support held fixed across spot, 3- to 36-month and asset denominators; separate threshold-only counterfactual (Section~\ref{sec:counterfactuals}). \\
Persistence & Repeated status switching and removals relate to prices and liquidity \citep{ChenNgo2017,ZainudinKhaw2024,AshrafEtAl2025}. & Reversal rates by driver, sustained reversals, and the reversal gap conditional on breach severity (Section~\ref{sec:counterfactuals}). \\
Continuous measures & Compliance degrees and rankings for single markets or rulebooks \citep{OrhanIsiker2021,AlnamlahEtAl2022,HameedMuneeza2024}. & Multi-rulebook, threshold-anchored score, tested for incremental breach information; it adds little beyond the target rule's own margin (Section~\ref{sec:counterfactuals}). \\
Portfolio implications & Screens change the investable universe, exposures and costs \citep{BoudtRazaWauters2019,Raza2021,PedersenEtAl2021}. & Flexible multi-standard objectives separated from binding single-rule mandates; emulated breaches compared with ETF and active-manager holdings (Section~\ref{sec:portfolios}). \\
\bottomrule
\end{tabularx}
\end{table}

Four strands of work are closest. Comparative studies show that screening rules produce inconsistent universes because their interpretations, components, thresholds and denominators differ \citep{KhatkhatayNisar2007,DerigsMarzban2008,AbdulRahmanEtAl2010,Ho2015,RizaldyAhmed2019}, and common-universe portfolio studies compare the resulting portfolios \citep{AshrafKhawaja2016,Raza2021}. Asset- and market-capitalisation denominators select different firms \citep{MarzbanAsutay2012}, and prior work warns that price volatility in a market-capitalisation denominator can destabilise status and motivates averaging \citep{AyedhEtAl2019,ParlakEtAl2024}. A second strand replaces binary labels with degrees, scores or rankings of compliance \citep{OrhanIsiker2021,AlnamlahEtAl2022,HameedMuneeza2024}, and recent qualitative work advocates reform of current screens \citep{MalikEtAl2025}. A third shows that index revisions, status removals and repeated switching relate to prices, liquidity and firm outcomes \citep{ChenNgo2017,KassimEtAl2017,ZainudinKhaw2024,AshrafEtAl2025}, and that screening affects firm financing decisions \citep{SukorHalim2022}. A fourth, in sustainable finance more broadly, shows that rating disagreement arises from scope, measurement and aggregation \citep{BergEtAl2022,ChristensenEtAl2022} and affects portfolios and returns \citep{GibsonBrandonEtAl2021,AvramovEtAl2022,PedersenEtAl2021}. Shariah rulebooks are explicit enough to connect disagreement and instability to observable threshold and denominator designs, which ESG ratings rarely allow.

\section{Data, screen reconstruction and CSCI}\label{sec:data}

\subsection{Data and screen reconstruction}\label{sec:datarec}

The monthly panel links CRSP securities to Compustat from January 1999 to December 2024 and contains 1,342,606 security-months. Accounting data are annual Compustat statements. Each statement enters the panel with a fixed lag, in the sixth month after its fiscal year-end month: a December 2010 year-end is first used in June 2011. Actual filing dates are not used; the fixed lag exceeds the 60 to 90 days that U.S. issuers have to file an annual report. A statement remains in force until the next annual statement enters, and for at most 18 months after its own entry. Security-months with no statement in force are dropped, and items missing from a statement stay missing rather than being filled. The cap rarely binds: 99.86\% of security-months use a statement that entered within the previous 12 months, and 56 security-months, under 0.01\%, carry an older statement than the cap allows because of irregular reporting histories in the source panel. Section~\ref{sec:counterfactuals} reports results with a one-month longer lag and with a 12-month cap. Market capitalisation equals absolute CRSP price times CRSP shares outstanding; trailing 3-, 6-, 12-, 24- and 36-month denominators require complete windows. Debt, cash and interest-bearing assets, receivables, total assets and sales follow a documented Compustat mapping. The non-permissible-income proxy is $\max(\mathrm{IDIT},0)+\max(\mathrm{NOPI},0)$ divided by sales. Non-operating income includes items, such as disposal gains, that no standard treats as impermissible, so the proxy is not equivalent to provider-classified non-permissible revenue; every result that uses it is therefore an approximation, and we show how the main results change without it. Business-sector permissibility uses SIC codes.

All screen states are reconstructed from uncapped ratios, and missing inputs remain missing rather than becoming failures. Where a rulebook changed, the old specification applies through the month before the effective change (August 2023 for DJIM and S\&P, July 2010 for MSCI's main series) and the new specification from the effective month; transition pairs that straddle a change are excluded. Cross-standard comparisons require all seven states and a permissible sector, which gives 939,826 observations. Transition pairs require consecutive months, observed states at both ends and a permissible sector at the origin month. Portfolios use prior-month information to earn month-$t$ CRSP total returns, which include dividends. Table~\ref{tab:rulebooks} lists the emulated rules, the months in which each specification applies, and the date from which each rulebook was in force.

\begin{table}[!htbp]
\centering
\caption{Emulated financial-ratio rulebooks}\label{tab:rulebooks}
\setlength{\tabcolsep}{4.5pt}
\begin{tabular}{lllll ll}
\toprule
 & & \multicolumn{3}{c}{Thresholds (\%)} & & \\
\cmidrule(lr){3-5}
Rulebook & Denominator & Debt & Cash & Receivables & Applied & In force \\
\midrule
AAOIFI & Spot market cap & 30 & 30 & -- & 1999--2024 & May 2004 \\
DJIM (old) & 24m avg.\ market cap & 33 & 33 & 33 (R) & to Aug.\ 2023 & Feb.\ 1999 \\
DJIM (new) & 24m avg.\ market cap & 33 & -- & -- & from Sep.\ 2023 & Sep.\ 2023 \\
S\&P (old) & 36m avg.\ market cap & 33 & 33 & 49 (R) & to Aug.\ 2023 & Dec.\ 2006 \\
S\&P (new) & 36m avg.\ market cap & 33 & -- & -- & from Sep.\ 2023 & Sep.\ 2023 \\
MSCI M-series & 36m avg.\ market cap & 33.3 & 33.3 & 49 (C$+$R) & 1999--2024 & Sep.\ 2015 \\
FTSE & Total assets & 33.3 & 33.3 & 50 (C$+$R) & 1999--2024 & Jan.\ 2008 \\
MSCI main (old) & Total assets & 33.3 & 33.3 & 70 (R) & to Jul.\ 2010 & Jul.\ 2007 \\
MSCI main (new) & Total assets & 33.3 & 33.3 & 33.3 (C$+$R) & from Aug.\ 2010 & Aug.\ 2010 \\
SC Malaysia & Total assets & 33 & 33 & -- & 1999--2024 & Nov.\ 2013 \\
\bottomrule
\end{tabular}

\tabnotes{Each threshold is a maximum ratio of the item to the rulebook's denominator. Cash includes interest-bearing assets. R denotes receivables and C$+$R cash plus receivables. Every rulebook also applies the business-sector gate and, in the binary emulations only, the approximate non-permissible-income test (proxy/sales $\leq5\%$). Before the in-force date a rulebook is applied retrospectively; Section~\ref{sec:disagreement} reports results using only rulebooks in force. Changes follow the S\&P Dow Jones Indices announcements of August 2023 (effective September 2023) and the MSCI methodologies of September 2008 and August 2010; switching MSCI in October 2008 instead raises cross-standard disagreement from 35.9\% to 36.2\%. Entry thresholds and buffers are not emulated. The SC's benchmarks count conventional debt and cash only; the U.S. data cannot separate Islamic financing, so SC Malaysia is applied to total debt and cash.}
\end{table}

\subsection{The continuous measure}\label{sec:csci}

Let $r_{ikts}\geq0$ be firm $i$'s ratio for component $k$ at time $t$ under standard $s$, with threshold $\theta_{ks}>0$. The binary decision $B_{ikts}=\mathbf{1}\{r_{ikts}\leq\theta_{ks}\}$ discards the signed margin $(\theta_{ks}-r_{ikts})/\theta_{ks}$: a firm 1\% inside the boundary and one 90\% inside receive the same label. We map each ratio to
\begin{equation}
c_{ikts}=\frac{\theta_{ks}}{\theta_{ks}+r_{ikts}},
\end{equation}
which is smooth, decreasing, bounded in $(0,1]$, equals 0.5 at the threshold and has no estimated parameter. Standard-level financial compliance is the mean over a standard's financial components, $C_{its}=K_s^{-1}\sum_k c_{ikts}$, and $\CSCI_{it}=S^{-1}\sum_{s}C_{its}$ averages the seven standards. All standard scores must be observed. Equal weights make the aggregation transparent, not uniquely correct, and the appendix tests geometric, harmonic and minimum aggregation, provider-family weights, leave-one-out scores and alternative mappings. The primary score excludes the business-sector gate and the income proxy, and it imposes no compliance threshold of its own: each standard retains its binary decision. A higher score is not a higher degree of permissibility. A security is either eligible under a governing rule or it is not; CSCI measures how far its balance sheet sits from each rule's boundaries.

The interpretation is a mechanism, not an equilibrium model. Screened investors with heterogeneous admissibility rules and unconstrained conventional investors hold the same securities. Shariah-compliant stocks therefore have a wide investor base but a narrow investment base for constrained investors, and conventional trading can absorb or swamp any demand effect of a change in compliance state \citep{Merton1987}. What constrained investors cannot avoid is the turnover that a binary boundary imposes on them. We therefore use the framework to motivate measurement and implementation, not to predict that CSCI is priced.

\section{Classification disagreement}\label{sec:disagreement}

\begin{figure}[!htbp]
\centering\includegraphics[width=.80\textwidth]{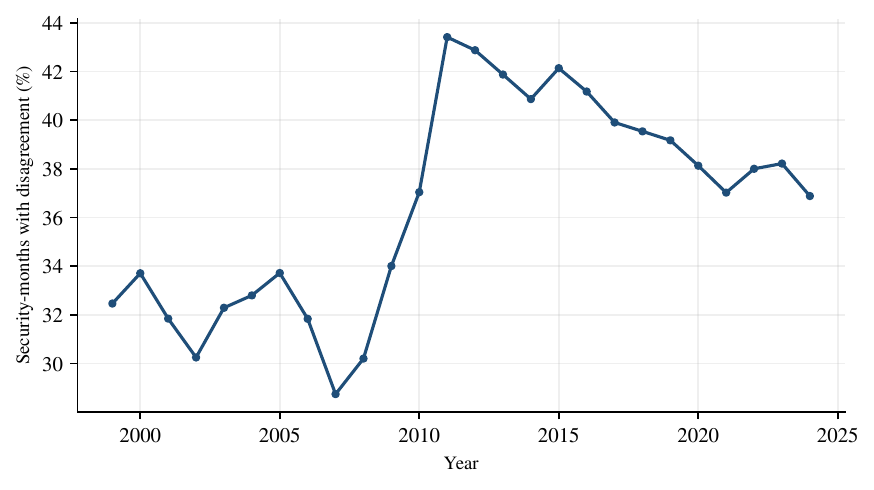}
\caption{Cross-standard disagreement over time. The sample requires permissible sectors and complete inputs for all seven rulebooks; all formulas are applied retrospectively.}
\label{fig:disagreement}
\end{figure}

Three disagreement statistics answer different questions, and we report all of them. The headline statistic applies all seven current formulas retrospectively to every month. Across the 939,826 complete permissible-sector observations, all standards fail 46.3\%, all pass 17.9\%, and the remaining 35.9\% exhibit disagreement. The largest pairwise disagreements, about 25--27\%, are between market-capitalisation and asset-based specifications. Disagreement varies over time (Figure~\ref{fig:disagreement}): it ranges from 29\% to 34\% between 1999 and 2008, peaks at 43\% in 2011 and declines to 37\% by 2024. Part of the rise coincides with MSCI's stricter cash-plus-receivables limit from August 2010, although it begins in 2009. The second statistic restricts the sample to September 2015 onwards, when all seven rulebooks were in force, so that retrospective and historical designs coincide; disagreement is then 38.8\%. The third uses, in every month, only the rulebooks in force at that date, requiring at least two; from May 2004 it is 32.7\%, lower mainly because the early years compare two to five rulebooks rather than seven. Dropping the SC Malaysia screen, which is retrospective before November 2013, lowers the headline figure to 34.6\%.

The income proxy matters for these levels. Without the approximate income test, all-fail observations fall from 46.3\% to 33.1\% and disagreement rises to 47.3\%, because the proxy fails many securities under every rulebook at once. The two figures answer different questions. The proxy-inclusive 35.9\% describes the full binary emulations; the financial-ratio 47.3\% is the cleaner measure of disagreement that comes from screen architecture, because it does not depend on a proxy that we regard as noisy. Either way, the rulebooks disagree on a large share of securities, and the financial-ratio screens disagree more, not less.

CSCI is continuous in practice, not only in construction. In permissible sectors its mean is 0.662 and its interquartile range 0.592--0.746 (Table~\ref{tab:csci}); 88.9\% of observations lie between 0.1 and 0.8 and 23.9\% between 0.4 and 0.6, and none lies below 0.1. The composite does depend on denominator architecture. Scores from the four market-capitalisation rulebooks and the three asset-based rulebooks have a mean monthly rank correlation of only 0.32, and CSCI tracks the market-capitalisation block (0.92) more closely than the asset block (0.64) because four of the seven rulebooks use market capitalisation. This is not a robustness detail but the paper's subject: the same firms rank differently depending on the denominator. Equal provider-family weights, which reduce the weight of the related DJIM and S\&P rules, give a score with a rank correlation of 0.994 with the baseline, and alternative mappings and aggregations are reported in Table~\ref{tab:aggregation}. A CSCI built only from rulebooks in force has a mean monthly rank correlation of 0.985 with the primary score.

\begin{table}[!htbp]
\centering
\caption{Distribution of the primary CSCI}\label{tab:csci}\begin{tabular}{lrrrrrr}
\toprule
Sample & $N$ & Mean & SD & P25 & Median & P75 \\
\midrule
Full universe & 1,197,410 & 0.617 & 0.156 & 0.518 & 0.647 & 0.730 \\
Permissible sectors & 965,162 & 0.662 & 0.121 & 0.592 & 0.674 & 0.746 \\
Eligible under at least one standard & 504,822 & 0.715 & 0.100 & 0.660 & 0.723 & 0.781 \\
\bottomrule
\end{tabular}

\tabnotes{The primary CSCI averages financial-component scores and excludes the sector gate and the approximate income ratio.}\end{table}

\section{Why eligibility changes}\label{sec:drivers}

Consecutive binary states form $C\to C$, $C\to NC$, $NC\to C$ and $NC\to NC$. Table~\ref{tab:transitions} reports transition probabilities for securities in permissible sectors at the origin month. AAOIFI is the least persistent: 4.60\% of compliant states breach the next month and 2.61\% of non-compliant states recover, and AAOIFI changes state in 3.36\% of all consecutive pairs. This comparison is descriptive because standards differ in components and thresholds as well as denominators; Section~\ref{sec:counterfactuals} holds those fixed.

\begin{table}[!htbp]
\centering
\caption{Monthly transition frequencies by methodology}\label{tab:transitions}\fittable{\begin{tabular}{lrrrrrr}
\toprule
Standard & Pairs & Any (\%) & $C\to NC$ & $p$ (\%) & $NC\to C$ & $p$ (\%) \\
\midrule
AAOIFI & 1,031,824 & 3.36 & 17,941 & 4.60 & 16,713 & 2.61 \\
DJIM & 954,684 & 1.63 & 8,184 & 2.26 & 7,330 & 1.24 \\
S\&P & 926,731 & 1.56 & 7,570 & 2.03 & 6,875 & 1.24 \\
FTSE & 1,031,856 & 1.13 & 5,829 & 1.71 & 5,780 & 0.84 \\
MSCI main & 1,028,801 & 1.12 & 5,818 & 1.82 & 5,711 & 0.81 \\
MSCI M-series & 929,487 & 1.54 & 7,511 & 2.11 & 6,849 & 1.19 \\
SC Malaysia & 1,031,856 & 1.19 & 6,308 & 1.62 & 5,982 & 0.93 \\
\bottomrule
\end{tabular}
}
\tabnotes{Pairs require consecutive months, observed states and a permissible sector at the origin month. Pairs that straddle a rulebook change are excluded. Direction-specific probabilities condition on the origin state.}\end{table}

For a ratio $R_t=N_t/D_t$, the counterfactuals $R^{D}_{t+1}=N_t/D_{t+1}$ and $R^{N}_{t+1}=N_{t+1}/D_t$ change one input at a time. A change is denominator only when $R^D$ crosses the threshold but $R^N$ does not, and numerator only in the converse case; ``both'' means either change is sufficient, ``joint'' that only their combination crosses, and ``mixed'' that several components change. These are arithmetic sufficiency statements, not causal decompositions. Two definitions recur below and should not be confused. The 65.6\% figure is the share of AAOIFI breaches that are denominator only with market capitalisation as the denominator, so it counts any movement in market capitalisation, whether from price or shares. The 63.5\% figure in Table~\ref{tab:priceshares} is the share in which the price change alone, with shares held at their earlier value, is sufficient.

For AAOIFI, denominator-only movement accounts for 66.1\% of $C\to NC$ and 75.2\% of $NC\to C$ transitions (Table~\ref{tab:drivers}). Market-capitalisation movement alone accounts for 65.6\% and 72.9\%; the remaining denominator-only events are movements in sales, the denominator of the income test. Numerator-only movement accounts for 27.2\% and 16.2\%. Excluding the 8.2\% of breaches and 10.5\% of recoveries in which only the income test crosses, market-capitalisation movement alone is sufficient for 71.4\% and 81.5\%. The pattern is specific to the denominator architecture: numerator-only shares of breaches are 78.8\% for FTSE, 79.3\% and 80.6\% for the two MSCI main specifications, and 80.7\% for SC Malaysia, with trailing-average market-capitalisation standards in between.

\begin{table}[!htbp]
\centering
\caption{Transition-driver sufficiency (percent within direction)}\label{tab:drivers}\begin{tabular}{lrrrrcrrrr}
\toprule
 & \multicolumn{4}{c}{$C\to NC$} & & \multicolumn{4}{c}{$NC\to C$} \\
\cmidrule(lr){2-5}\cmidrule(lr){7-10}
Standard & Denom. & Numer. & Joint & Both & & Denom. & Numer. & Joint & Both \\
\midrule
AAOIFI & 66.1 & 27.2 & 2.9 & 2.3 & & 75.2 & 16.2 & 3.0 & 4.2 \\
DJIM (old) & 32.7 & 61.4 & 2.4 & 1.7 & & 47.3 & 40.4 & 4.1 & 6.0 \\
DJIM (new) & 23.4 & 65.9 & 4.8 & 4.1 & & 44.0 & 44.0 & 5.0 & 5.7 \\
S\&P (old) & 25.3 & 69.0 & 2.1 & 1.8 & & 41.4 & 46.9 & 3.9 & 5.4 \\
S\&P (new) & 18.5 & 70.6 & 5.3 & 4.1 & & 40.6 & 49.0 & 3.9 & 5.8 \\
MSCI M-series & 26.6 & 67.2 & 2.4 & 1.7 & & 43.9 & 44.4 & 3.8 & 5.4 \\
\addlinespace
FTSE & 8.1 & 78.8 & 4.9 & 3.0 & & 18.5 & 50.7 & 6.4 & 13.2 \\
MSCI main (old) & 7.6 & 79.3 & 5.5 & 3.6 & & 15.8 & 54.2 & 7.8 & 15.7 \\
MSCI main (new) & 7.4 & 80.6 & 4.3 & 3.6 & & 19.9 & 51.3 & 8.8 & 11.5 \\
SC Malaysia & 6.7 & 80.7 & 5.2 & 3.3 & & 15.3 & 55.2 & 8.6 & 15.2 \\
\bottomrule
\end{tabular}

\tabnotes{Percent of state changes in each direction. Denom.\ means the denominator change alone is sufficient, Numer.\ the numerator change alone; Joint and Both are defined in the text. Omitted shares are mixed-component and unresolved events. Denom.\ includes movements in sales, the denominator of the approximate income test. Old and new DJIM, S\&P and MSCI main specifications are restricted to their applicable periods. Market-capitalisation rulebooks are listed first, asset-based rulebooks after the gap.}\end{table}

For market-capitalisation screens we next separate $M=PS$, using $M^{P}_{t+1}=P_{t+1}S_t$ and $M^{S}_{t+1}=P_tS_{t+1}$ with the accounting numerator fixed. Price movement alone is sufficient for 63.5\% of AAOIFI breaches and 67.7\% of recoveries; shares alone account for 0.6\% and 2.2\% (Table~\ref{tab:priceshares}). Most importantly, 7,036 breaches (39.2\% of all AAOIFI breaches) and 6,875 recoveries (41.1\%) occur with an unchanged accounting vintage, unchanged accounting values, unchanged shares and an unchanged sector gate, and every one is reproduced by the price counterfactual. Only these events are described as cleanly price-driven. \citet{Rai2026} attributes 54.0\% of AAOIFI leverage-screen flips in a quarterly S\&P 500 panel to price alone. His design uses one screen, quarterly sampling of current constituents and a different counterfactual classification, so the shares are not directly comparable, but both designs find that price movement, not balance-sheet change, accounts for most flips. The AAOIFI results do not depend on retrospective application: from June 2004, after the standard was adopted, denominator-only shares are 66.2\% and 75.3\%.

Nor do they depend on the income proxy. Applying only AAOIFI's two financial-ratio screens, securities change state in 4.41\% of consecutive permissible-sector pairs, against 3.36\% with the proxy, because the proxy no longer masks ratio changes in firms that fail the income test. Market-capitalisation movement alone is sufficient for 71.6\% of these breaches and 81.6\% of recoveries, and 49.0\% of the denominator-only breaches reverse within three months, against 15.4\% of numerator-only breaches.

\begin{table}[!htbp]
\centering
\caption{Price-versus-shares decomposition of AAOIFI transitions}\label{tab:priceshares}\fittable{\begin{tabular}{lrr}
\toprule
Sufficiency category & $C\to NC$: N (\%) & $NC\to C$: N (\%) \\
\midrule
Price only & 11,391 (63.5) & 11,309 (67.7) \\
Shares only & 116 (0.6) & 362 (2.2) \\
Price and shares (each sufficient) & 150 (0.8) & 299 (1.8) \\
Price--shares joint & 120 (0.7) & 228 (1.4) \\
Numerator only & 3,718 (20.7) & 2,031 (12.2) \\
Numerator--denominator mixed & 879 (4.9) & 679 (4.1) \\
Multiple binding components & 60 (0.3) & 46 (0.3) \\
Other / unresolved & 1,507 (8.4) & 1,759 (10.5) \\
\bottomrule
\end{tabular}
}
\tabnotes{Percentages use all direction-specific AAOIFI state changes of securities in permissible sectors at the origin month. ``Other / unresolved'' includes transitions caused by the income test, which has no market-capitalisation denominator, and events with missing price or share data. Categories are mutually exclusive at event level.}\end{table}

\section{Architecture counterfactuals and reversals}\label{sec:counterfactuals}

\subsection{Denominator experiments}\label{sec:denexp}

\begin{figure}[!htbp]
\centering\includegraphics[width=.72\textwidth]{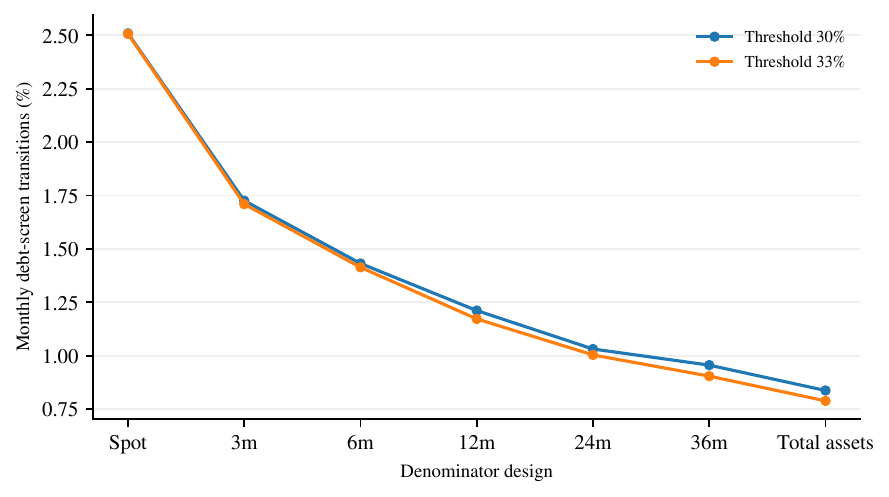}
\caption{Debt-screen transition frequency under alternative denominators on the common sample of Table~\ref{tab:denominators}, with the threshold fixed within each curve.}\label{fig:smoothing}
\end{figure}

Holding the debt threshold at 30\% and using the same 753,109 consecutive security-month pairs for every design, progressively smoothing market capitalisation reduces the monthly transition frequency monotonically: 2.51\% at spot, 1.73\% at three months, 1.43\% at six, 1.21\% at 12, 1.03\% at 24 and 0.96\% at 36 months; a total-assets counterfactual gives 0.84\% (Table~\ref{tab:denominators}, Figure~\ref{fig:smoothing}). On the same pairs, changing the spot threshold alone from 30\% to 33\% leaves the transition frequency unchanged at 2.51\% while raising the pass rate from 64.8\% to 66.9\% (Figure~\ref{fig:smoothing}). On all 1,076,801 pairs with a spot denominator, which are not restricted to common support, the corresponding frequencies are 2.60\% and 2.59\%. These are design experiments, not statements about what AAOIFI uses.

\begin{table}[!htbp]
\centering
\caption{Debt-screen denominator experiments at a fixed 30\% threshold}\label{tab:denominators}\begin{tabular}{lrrrr}
\toprule
Denominator & Pairs & Pass (\%) & Transitions & Frequency (\%) \\
\midrule
Spot & 753,109 & 64.83 & 18,894 & 2.51 \\
3-month average & 753,109 & 64.83 & 12,996 & 1.73 \\
6-month average & 753,109 & 64.83 & 10,783 & 1.43 \\
12-month average & 753,109 & 64.90 & 9,123 & 1.21 \\
24-month average & 753,109 & 65.03 & 7,768 & 1.03 \\
36-month average & 753,109 & 65.09 & 7,200 & 0.96 \\
Total assets & 753,109 & 67.21 & 6,306 & 0.84 \\
\bottomrule
\end{tabular}

\tabnotes{Every row uses the intersection of pairs with complete spot, 3-, 6-, 12-, 24- and 36-month market-capitalisation denominators and total assets. Total assets is a counterfactual design.}\end{table}

The mechanism is simple. With a fixed numerator $N>0$, $R=N/M$ breaches $\theta$ exactly when $M<N/\theta$. If $\log M_{t+1}$ given current information is normal with mean $\mu_t$ and standard deviation $\sigma$, the breach probability is $\Phi((\log(N/\theta)-\mu_t)/\sigma)$. For a security whose conditional mean lies on the compliant side of the boundary ($\mu_t>\log(N/\theta)$), this probability increases with denominator volatility $\sigma$; for a security whose conditional mean lies in breach, it decreases, so higher volatility raises the chance of recovery instead. A $w$-month average has lower variance than spot market capitalisation unless observations are perfectly correlated. Lower variance does not imply fewer state changes for every process, so the appendix reports a simulation rather than a theorem.

\subsection{Reversals and breach severity}\label{sec:reversals}

Denominator-driven changes also reverse faster. A reversal occurs when a new state returns to its origin within the horizon. Of AAOIFI breaches driven by the denominator alone, 49.7\% reverse within three months, against 11.1\% of numerator-only breaches; the gap of 38.5 percentage points has a $t$-statistic of 25.1 with standard errors clustered by firm and month (Table~\ref{tab:reversals}). By 12 months the rates are 70.6\% and 35.8\%. The ordering survives when a reversal must last three consecutive months (53.1\% against 18.0\% within 12 months; Table~\ref{tab:sustained}) and when the sample starts in June 2004 (49.4\% against 11.9\%).

\begin{table}[!htbp]
\centering
\caption{AAOIFI short-horizon reversal probabilities (percent)}\label{tab:reversals}\begin{tabular}{llrrrr}
\toprule
Direction & Driver & 1m & 3m & 6m & 12m \\
\midrule
$C\to NC$ & denominator only & 29.5 & 49.7 & 61.4 & 70.6 \\
$C\to NC$ & numerator only & 4.9 & 11.1 & 17.6 & 35.8 \\
$NC\to C$ & denominator only & 27.5 & 46.9 & 58.8 & 69.8 \\
$NC\to C$ & numerator only & 7.8 & 19.1 & 27.9 & 49.3 \\
\bottomrule
\end{tabular}
\end{table}

Denominator-only breaches are, however, shallower. The median lands 7.8\% beyond the binding threshold, against 41.3\% for numerator-only breaches, and shallow breaches reverse more often whatever their cause. Table~\ref{tab:severity} therefore compares breaches of similar severity. Controlling for log severity reduces the three-month gap from 36.1 to 20.5 percentage points with transition-month fixed effects; severity-decile indicators reduce it to 16.9; adding log market capitalisation, prior return volatility and the age of the accounting record reduces it to 14.6 (standard error 1.6). Within every severity decile, denominator-only breaches reverse more often. Severity explains more than half of the raw gap, and the driver still carries information about persistence. The remaining gap mixes two things this design cannot separate: prices move continuously, while accounting numerators change only when a report arrives. The comparison is descriptive.

\begin{table}[!htbp]
\centering
\caption{AAOIFI three-month reversal gap conditional on breach severity}\label{tab:severity}\begin{tabular}{lrrr}
\toprule
Specification & Gap (pp) & s.e. & Events \\
\midrule
Transition-month fixed effects only & 36.1 & 1.4 & 16,585 \\
+ log severity & 20.5 & 1.3 & 16,585 \\
+ severity deciles & 16.9 & 1.3 & 16,585 \\
+ severity deciles, size, volatility, report age & 14.6 & 1.6 & 14,916 \\
\midrule
\multicolumn{4}{l}{\textit{Median severity (distance past the boundary, \% of threshold)}} \\
Denominator-only breaches & 7.8 & & 11,744 \\
Numerator-only breaches & 41.3 & & 4,841 \\
\bottomrule
\end{tabular}

\tabnotes{Linear probability models of a three-month reversal for AAOIFI $C\to NC$ events driven by the denominator alone or the numerator alone. The gap is the coefficient on the denominator-only indicator. All models include transition-month fixed effects; standard errors are clustered by firm and month. Severity is $\max_k r_k/\theta_k-1$ at the transition month; report age is months since the accounting record in force was dated.}\end{table}

\subsection{Does the continuous score add information?}\label{sec:information}

CSCI varies within every binary state. Among AAOIFI-compliant observations, next-month breach frequency falls from 9.7\% in the lowest CSCI decile to 1.0\% in the highest (Figure~\ref{fig:gradient}). In linear probability models with firm and month fixed effects, a one-standard-deviation higher CSCI is associated with a 3.39 percentage-point lower AAOIFI breach probability, and with lower breach probabilities under every standard (Table~\ref{tab:lpm}). These estimates are partly mechanical because the target rule's inputs enter CSCI. When CSCI excludes the target rulebook and the regression controls for the target's nearest component margin, boundary distance dominates: the target-excluded score remains negative for S\&P and the MSCI M-series, is indistinguishable from zero for DJIM, and becomes positive for AAOIFI and the asset-based standards (Table~\ref{tab:incremental}). CSCI preserves a descriptive ranking but is not a general breach predictor once the governing rule's own margin is known.

\begin{table}[!htbp]
\centering
\caption{Target-excluded CSCI conditional on the target boundary}\label{tab:incremental}\fittable{\begin{tabular}{lrrrrr}
\toprule
Standard & $N$ & $CSCI_{-s}$ (pp) & $t$ & Target margin (pp) & $t$ \\
\midrule
AAOIFI & 354,469 & 1.38 & 12.24 & -6.36 & -28.34 \\
DJIM & 353,956 & 0.05 & 0.83 & -2.14 & -22.90 \\
S\&P & 372,427 & -0.27 & -3.75 & -1.49 & -18.76 \\
FTSE & 322,406 & 0.14 & 2.33 & -0.72 & -8.90 \\
MSCI main & 300,521 & 0.15 & 2.50 & -0.74 & -7.92 \\
MSCI M-series & 355,481 & -0.26 & -4.82 & -1.58 & -18.82 \\
SC Malaysia & 364,041 & 0.17 & 2.66 & -0.64 & -8.48 \\
\bottomrule
\end{tabular}
}
\tabnotes{The outcome is a next-month $C\to NC$ transition. Both standardised regressors enter jointly, with firm and month fixed effects and two-way clustered standard errors.}\end{table}

\subsection{Robustness}

Firm-level and timing variants leave these results nearly unchanged. Selecting one primary security per firm gives 35.87\% disagreement, a 66.07\% denominator-only share of AAOIFI breaches and a 49.66\% three-month reversal rate. Delaying the entry of each statement by a further month, or capping its life at 12 rather than 18 months after entry, changes none of these by more than 0.6 percentage points (Table~\ref{tab:firmrobust}).

\section{Portfolio and holdings implications}\label{sec:portfolios}

\subsection{Flexible multi-standard objectives}\label{sec:flexible}

We compare the market, binary AAOIFI, the union of the seven binary screens, and CSCI tilts with weights proportional to $\mathrm{ME}_{i,t-1}\CSCI_{i,t-1}^{\kappa}$ over permissible-sector securities; $\kappa=0$ is the cap-weighted permissible-sector portfolio. All use one-month-lagged information. Table~\ref{tab:portfolios} reports gross and net performance. The linear and squared tilts trade 9.0\% and 11.8\% a year, against 41.2\% for binary AAOIFI and 23.2\% for the union. Their Sharpe ratios are 0.493 and 0.484, against 0.523 for binary AAOIFI, and none of the screened portfolios or CSCI tilts in the table has a significant six-factor alpha; their alphas lie within $\pm0.23\%$ a year. The market portfolio's alpha of $-0.13\%$ a year is statistically significant ($t=-3.51$) but economically negligible; it reflects small differences between the panel's CRSP universe and the market factor. Tilts built only from the asset-based rulebooks trade as little but have lower Sharpe ratios (0.462 and 0.450), and the squared asset-based tilt has a negative alpha of $-0.86\%$ a year ($t=-2.05$; Table~\ref{tab:blocks}). Net of 50 basis points per round trip, the binary AAOIFI Sharpe ratio falls to 0.510 and the tilts' to 0.490 and 0.480. Compounding CRSP delisting returns into the delisting month changes no Sharpe ratio by more than 0.0004 and no alpha by more than 0.02 percentage points. Against continuous and buffered competitors at the same weighted share of standards passed, the linear tilt trades 9.0\% where the least-trading AAOIFI-boundary tilt trades 12.5\% (Figure~\ref{fig:frontier}); when the target is only the AAOIFI margin, the advantage of the squared tilt shrinks to 11.8\% against 12.5\%.

\begin{table}[!htbp]
\centering
\caption{Lagged-weight portfolio performance}\label{tab:portfolios}\fittable{\begin{tabular}{lrrrrrrrr}
\toprule
Strategy & Excess ret. & Vol. & Sharpe & Alpha & $t(\alpha)$ & Turnover & Net Sharpe & Net alpha \\
\midrule
Market & 7.41 & 15.86 & 0.467 & -0.13 & -3.51 & 6.4 & 0.465 & -0.16 \\
Binary AAOIFI & 8.12 & 15.53 & 0.523 & -0.06 & -0.12 & 41.2 & 0.510 & -0.26 \\
Binary union & 8.48 & 15.75 & 0.539 & 0.14 & 0.33 & 23.2 & 0.531 & 0.02 \\
CSCI tilt ($\kappa=0$) & 7.87 & 15.76 & 0.500 & 0.05 & 0.18 & 6.5 & 0.498 & 0.01 \\
CSCI tilt ($\kappa=1$) & 7.70 & 15.63 & 0.493 & -0.08 & -0.27 & 9.0 & 0.490 & -0.12 \\
CSCI tilt ($\kappa=2$) & 7.52 & 15.54 & 0.484 & -0.23 & -0.70 & 11.8 & 0.480 & -0.29 \\
\bottomrule
\end{tabular}
}
\tabnotes{311 monthly returns. Excess return, volatility, alpha and turnover are annualised percentages. Alpha uses market, size, value, profitability, investment and momentum factors with HAC(6) errors. Turnover is one-way and drift-adjusted, one-half the sum of absolute weight changes. Net columns deduct 50 basis points per round trip applied to one-way turnover.}\end{table}

\subsection{Binding single-rule mandates}\label{sec:binding}

The tilts above may hold securities that fail AAOIFI, which a fund governed by AAOIFI cannot do. Table~\ref{tab:hardmandate} restricts every strategy to the same set, securities AAOIFI-compliant at $t-1$, and splits turnover into the weight sold because a holding became non-compliant (breach exits), other exits, and discretionary trading. The contrast of Table~\ref{tab:portfolios} disappears. Cap weighting trades 40.8\% a year, of which breach exits are 24.8 points. CSCI tilts cannot avoid forced exits, but they hold less in securities about to breach: turnover falls to 38.9\% ($\kappa=1$) and 38.0\% ($\kappa=2$). For $\kappa=2$ the paired reduction is 2.8 percentage points (95\% interval 1.6 to 4.2, moving-block bootstrap), worth about 1.4 basis points a year at 50 basis points per round trip. A tilt toward the AAOIFI margin cuts breach exits further, to 11.2\% at $\kappa=4$, but raises turnover to 51.2\% because it trades on the margin itself. For a binding single-rule mandate, CSCI offers a modest reduction in compliance-driven trading, not the large reduction available to a flexible objective.

\begin{table}[!htbp]
\centering
\caption{Portfolios bound by the same AAOIFI eligibility set}\label{tab:hardmandate}\setlength{\tabcolsep}{4pt}
\begin{tabular}{lrrrrrrrrr}
\toprule
 & \multicolumn{4}{c}{Turnover and exits (\% a year)} & & \multicolumn{2}{c}{Weighted score} & & \\
\cmidrule(lr){2-5}\cmidrule(lr){7-8}
Weighting & Total & Breach & Other & Discr. & Eff.\ $N$ & CSCI & Margin & Sharpe & Alpha ($t$) \\
\midrule
Cap weight & 40.8 & 24.8 & 2.0 & 14.0 & 66 & 0.741 & 0.653 & 0.538 & $-$0.03 ($-$0.06) \\
Equal weight & 108.2 & 47.4 & 6.4 & 54.3 & 1140 & 0.748 & 0.670 & 0.527 & 1.76 (1.77) \\
CSCI tilt, $\kappa=1$ & 38.9 & 23.2 & 2.0 & 13.7 & 66 & 0.747 & 0.657 & 0.528 & $-$0.19 ($-$0.38) \\
CSCI tilt, $\kappa=2$ & 38.0 & 21.8 & 2.0 & 14.3 & 66 & 0.753 & 0.661 & 0.519 & $-$0.34 ($-$0.71) \\
CSCI tilt, $\kappa=4$ & 38.1 & 19.3 & 1.9 & 16.8 & 63 & 0.765 & 0.670 & 0.499 & $-$0.65 ($-$1.33) \\
Margin tilt, $\kappa=1$ & 39.2 & 20.7 & 2.0 & 16.5 & 69 & 0.745 & 0.669 & 0.531 & $-$0.06 ($-$0.12) \\
Margin tilt, $\kappa=2$ & 42.1 & 17.0 & 2.0 & 23.1 & 70 & 0.750 & 0.686 & 0.522 & $-$0.13 ($-$0.25) \\
Margin tilt, $\kappa=4$ & 51.2 & 11.2 & 2.0 & 38.0 & 70 & 0.761 & 0.719 & 0.499 & $-$0.38 ($-$0.67) \\
\bottomrule
\end{tabular}

\tabnotes{Every strategy holds only securities that are AAOIFI-compliant at $t-1$ with observed CSCI and AAOIFI margin (the minimum of the continuous AAOIFI components), weighted by lagged market capitalisation times the signal raised to $\kappa$, except equal weight. Breach exits are the weight sold because a holding became non-compliant, other exits are holdings that leave the panel or lose inputs, and Discr.\ is the remaining discretionary turnover. Margin tilts weight by the AAOIFI margin. Turnover components are annualised percentages. Sharpe ratios and six-factor alphas (percent a year, HAC(6) $t$-statistics) are before costs; 311 months.}\end{table}

\FloatBarrier
\subsection{Fund holdings}\label{sec:holdings}

Do emulated breaches correspond to real decisions? Using quarter-end SEC Form N-PORT holdings of two index ETFs from May 2020 to December 2024, SPUS (S\&P rules) and HLAL (FTSE rules), a test specified before holdings were linked to emulated states finds that an emulated breach of the fund's rule coincides with a higher probability of sale by the next quarter: 4.9\% against 1.5\% for SPUS (Fisher $p=0.066$) and 8.7\% against 3.5\% for HLAL ($p=0.014$). Only 7.7\% and 13.0\% of exits coincide with a breach. An exploratory check on Saturna Capital, manager of the Amana Funds, whose rule divides debt by a 12-month average of market capitalisation \citep{Saturna2019,Saturna2021}, uses quarterly 13F holdings from 2004 to 2024: breach-associated exit rates are 2.7\% against 0.4\% for material positions ($p=0.034$, from only 14 exits) and 8.8\% against 3.2\% for all positions, and 7.5--14.3\% of exits coincide with a breach. AAOIFI breaches that the fund's smoothed rule does not register coincide with more sales in raw rates, but this effect largely disappears once price declines are controlled (2.0 percentage points, standard error 2.8). Because breach and exit are observed over the same quarter, these are associations, not trading sequences, and the 13F combines Amana with other accounts, although 97.5\% of its value sits in securities the Amana funds hold. The appendix reports the full design and tables. The holdings evidence is limited correspondence between emulated screens and portfolio changes, not validation of the mechanism.

\subsection{Practical use}\label{sec:practical}

For a fund manager, the evidence suggests a simple procedure. Keep the governing rule's binary decision. Monitor each holding's margin to that rule and, when a breach occurs, classify it by its observed driver; in this sample about half of denominator-only breaches reversed within three months. This is a description of past breaches, not a forecast for individual holdings. Use CSCI, or the rule's own margin, to tilt weights away from holdings close to the boundary. Under a binding mandate this lowers forced selling modestly; across several standards, as in a multi-mandate platform or index family, it lowers turnover substantially. None of this changes which securities are permissible, and it gives no reason to prefer weakly compliant securities: a score is not a licence to choose the most lenient reading of a standard.

\section{Conclusion}\label{sec:conclusion}

Binary values-based screens are shaped by their arithmetic as much as by their values. In U.S. data, seven Shariah rulebooks disagree on a large share of securities; under a spot market-capitalisation denominator, most eligibility changes are price events, and half of the breaches caused by the denominator alone reverse within three months; and holding the threshold fixed, the denominator design determines how often eligibility changes. Among breaches of similar severity the reversal gap narrows but persists. CSCI organises this evidence on one scale and supports low-turnover tilts for flexible multi-standard objectives, but under a binding single-rule mandate its benefit is modest, it adds little beyond the governing rule's margin, and it earns no positive abnormal return.

\paragraph{Implications for values-based screening}\label{sec:implications} The arithmetic is not specific to Shariah screening. Any screen that divides a slowly changing quantity by a market value inherits the volatility of prices. The EU Climate Transition and Paris-Aligned Benchmarks, for example, measure greenhouse-gas intensity as emissions divided by enterprise value including cash, which contains the market capitalisation of equity, and they adjust that denominator for changes in the average enterprise value of index constituents \citep{EU2020Benchmarks}. A firm's measured intensity, like its Shariah debt ratio, can therefore change when its share price moves and its emissions do not. The evidence here suggests how large such effects can be when the boundary is binary.

For standard setters the results identify a trade-off, not a recommendation. Four design responses are available, and each changes the rule. Averaging the denominator removes most short-lived changes but delays recognition of persistent deterioration. An asset denominator removes price sensitivity but measures a different economic concept. Buffers and grace periods leave the formula intact but slow every change, including those driven by fundamentals. Continuous weighting leaves the formula and the binary state intact and lets portfolios respond in proportion to the distance from the boundary, but its benefit depends on the mandate. Because the driver of a breach can be identified when it occurs, a provider could also treat denominator-only breaches differently from breaches caused by a change in debt, cash or receivables. Whether any of these is acceptable is a governance question.

For investors the main implication is transparency. Index providers and fund managers already observe the inputs needed to report why a security entered or left a screened universe. Reporting the share of changes driven by prices rather than by changes in the measured accounting inputs would let investors judge whether turnover reflects their values or the arithmetic of the rule. For Shariah-governed investors, the binary decision remains the governing one: a security is permissible or it is not, and a continuous score does not make a weakly compliant security more or less permissible.

\paragraph{Limitations} The emulations approximate published rules and cannot reproduce providers' revenue classifications or Shariah-board judgements; the income test uses an accounting proxy, so results that rely on it are approximations. Several rulebooks are applied retrospectively before they existed. The sample covers U.S. firms whose debt is conventional; markets where firms issue sukuk and where Shariah-sensitive ownership is larger may behave differently. And the transition evidence describes arithmetic sufficiency, not causation.

\section*{Declarations}
\paragraph{Funding} This research did not receive any specific grant from funding agencies in the public, commercial, or not-for-profit sectors.

\paragraph{Competing interests} The authors declare that they have no known competing financial interests or personal relationships that could have appeared to influence the work reported in this paper.

\paragraph{Data availability} CRSP, Compustat and LSEG data are licensed and cannot be redistributed. SEC N-PORT filings and the screening methodologies are public. The analysis code, the aggregate outputs reported in the paper and a claims-validation script that checks every number in the text are available from the corresponding author and will be released publicly on journal publication; licensed and security-level data are excluded.

\paragraph{Use of generative AI} During the preparation of this work, the authors used Claude (Anthropic) and Codex (OpenAI) to solely assist with LaTeX debugging. All formulas, empirical claims, references, notation, and manuscript content were subsequently reviewed and verified by the authors, who take full responsibility for the content of the submitted manuscript.

\clearpage
\appendix
\setcounter{table}{0}\setcounter{figure}{0}\setcounter{equation}{0}
\renewcommand{\thetable}{A\arabic{table}}
\renewcommand{\thefigure}{A\arabic{figure}}
\renewcommand{\theequation}{A\arabic{equation}}
\begin{center}{\Large\bfseries Appendix\par}\end{center}
\vspace{4pt}
\noindent This appendix reports supporting analyses. Sections are lettered and tables and figures are numbered A1, A2, and so on.

\section{Replication classification and reproducibility}\label{sec:repl}
Official-emulation tables are confirmatory in the sense that thresholds and formulas are set by documented rules. The counterfactual driver, reversal, event-time, and daily analyses are exploratory and are labelled as such. Alternative mappings, denominator windows, thresholds, transaction costs, and daily sampling are robustness exercises. Every reported number maps to a generated CSV; the replication package records the sample flow, data dictionary and known limitations, and its validation script checks every number in the manuscript against the generated outputs.

\section{Analytical and simulation details}\label{sec:anal}
Let $\bar M_{t,w}=w^{-1}\sum_{j=0}^{w-1}M_{t-j}$. Stationarity gives
\begin{equation}
\operatorname{Var}(\bar M_{t,w})=\frac{\sigma_M^2}{w}\left[1+2\sum_{h=1}^{w-1}\left(1-\frac{h}{w}\right)\rho_h\right].
\end{equation}
For the illustration, 20,000 firms each have a fixed numerator $N=\theta$ and heterogeneous long-run log market capitalisation. Monthly log market capitalisation follows an AR(1) with $\rho=0.90$ for 312 months. At innovation standard deviation 0.10, simulated transition frequency declines from 7.81\% under spot market capitalisation to 1.06\% under a 24-month mean and 0.75\% under a 36-month mean. The same monotone ordering holds for innovation standard deviations 0.05 and 0.20. These values are not calibrated estimates; they demonstrate the smoothing mechanism under stated assumptions. Full results and the seed appear in the replication tables.

\section{Measurement robustness}\label{sec:measure}

\begin{figure}[!htbp]
\centering\includegraphics[width=.82\textwidth]{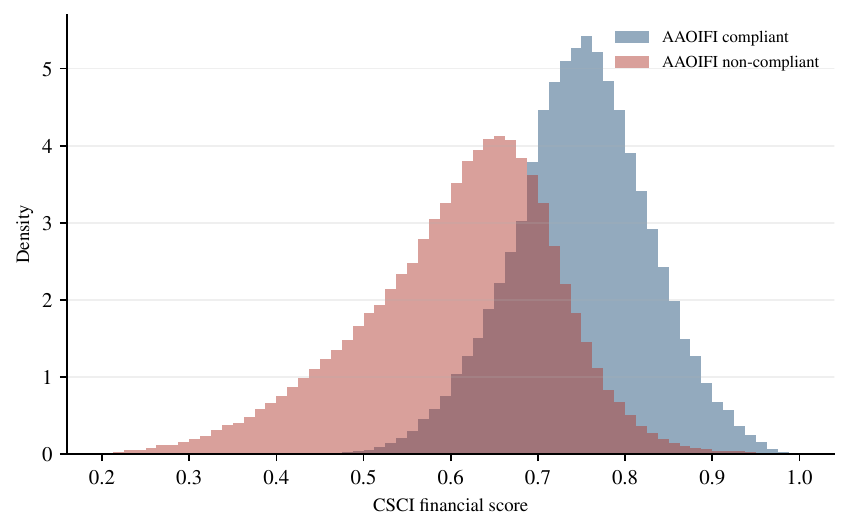}
\caption{CSCI distributions by emulated AAOIFI binary state, permissible sectors. Binary groups overlap because CSCI embeds multiple standards and retains within-state distance.}
\label{fig:status}
\end{figure}

The exponential and linear alternatives to the reciprocal mapping have rank correlations of 0.992 and 0.970 with the baseline. Table~\ref{tab:aggregation} addresses compensability. Geometric and harmonic component aggregation retain rank correlations of 0.983 and 0.952; the deliberately non-compensatory component minimum lowers this to 0.862. Equal provider-family weights yield 0.994. Leave-one-rulebook correlations range from 0.992 to 0.997, and excluding the entire S\&P/Dow Jones family gives the lowest family-exclusion correlation, 0.965. The broad ranking is not created by duplicating a provider family, although the minimum construction is meaningfully different. Figure~\ref{fig:status} shows that CSCI varies widely within both AAOIFI states.

\begin{table}[!htbp]
\centering
\caption{Aggregation and weighting robustness}\label{tab:aggregation}\fittable{\begin{tabular}{lrrrr}
\toprule
Construction & $N$ & Mean & SD & Spearman $\rho$ \\
\midrule
Baseline arithmetic & 1,197,410 & 0.617 & 0.156 & 1.000 \\
Geometric within rulebook & 1,197,410 & 0.582 & 0.163 & 0.983 \\
Harmonic within rulebook & 1,197,410 & 0.552 & 0.171 & 0.952 \\
Minimum within rulebook & 1,197,410 & 0.435 & 0.173 & 0.862 \\
Geometric across rulebooks & 1,197,410 & 0.601 & 0.174 & 0.998 \\
Harmonic across rulebooks & 1,197,410 & 0.586 & 0.189 & 0.994 \\
Minimum across rulebooks & 1,197,410 & 0.479 & 0.198 & 0.932 \\
Equal provider-family weights & 1,197,410 & 0.618 & 0.146 & 0.994 \\
\bottomrule
\end{tabular}
}
\tabnotes{Spearman correlations compare each construction with the baseline arithmetic score on common observations. Minimum aggregation is a non-compensatory stress test, not an alternative claim about the correct compliance rule.}\end{table}

\section{Additional transition and reversal evidence}\label{sec:trans}

\begin{figure}[!htbp]
\centering\includegraphics[width=.82\textwidth]{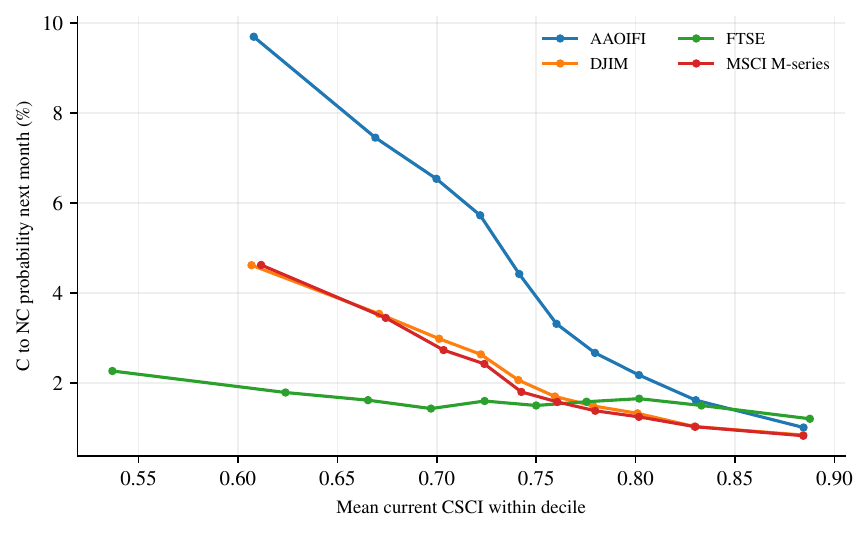}
\caption{Next-month $C\to NC$ frequency by within-standard CSCI decile for four standards. This is an in-sample information diagnostic, not a forecasting exercise.}\label{fig:gradient}
\end{figure}

\begin{figure}[!htbp]
\centering\includegraphics[width=.82\textwidth]{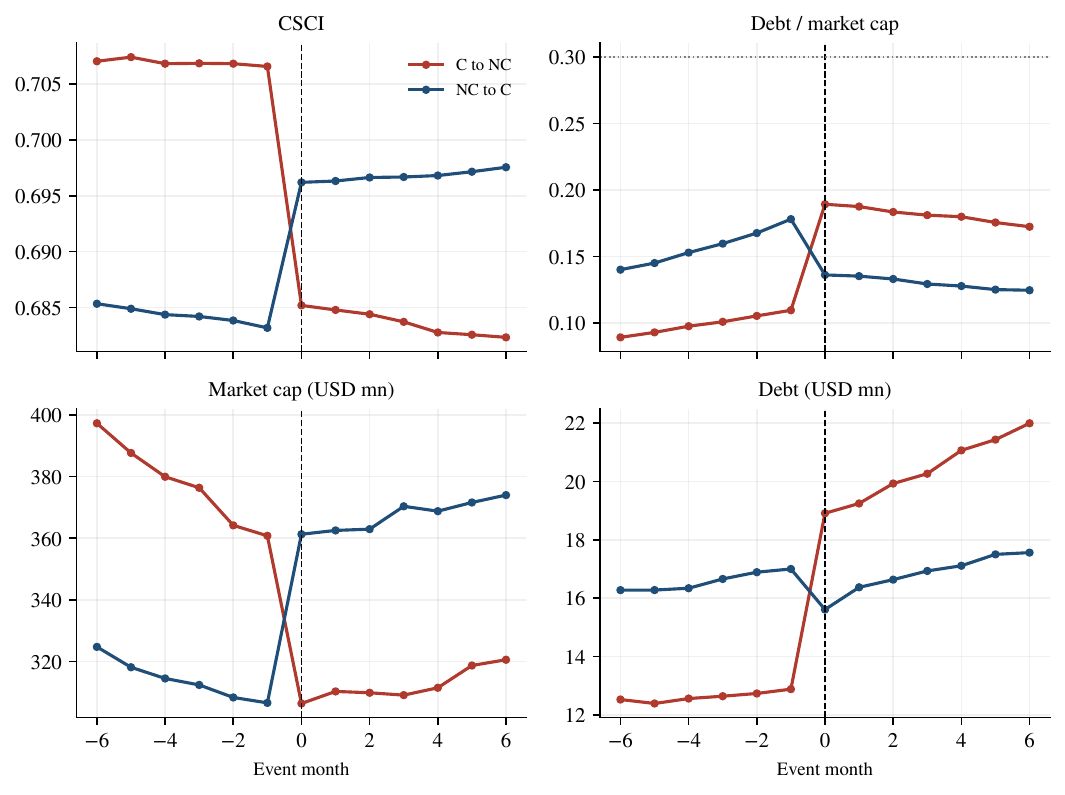}
\caption{AAOIFI state changes in event time. Lines are cross-event medians; month 0 is the first month in the new state. All observed changes of securities in permissible sectors at the origin month are included, so a security can contribute more than one event.}
\label{fig:event}
\end{figure}

\begin{figure}[!htbp]
\centering\includegraphics[width=.94\textwidth]{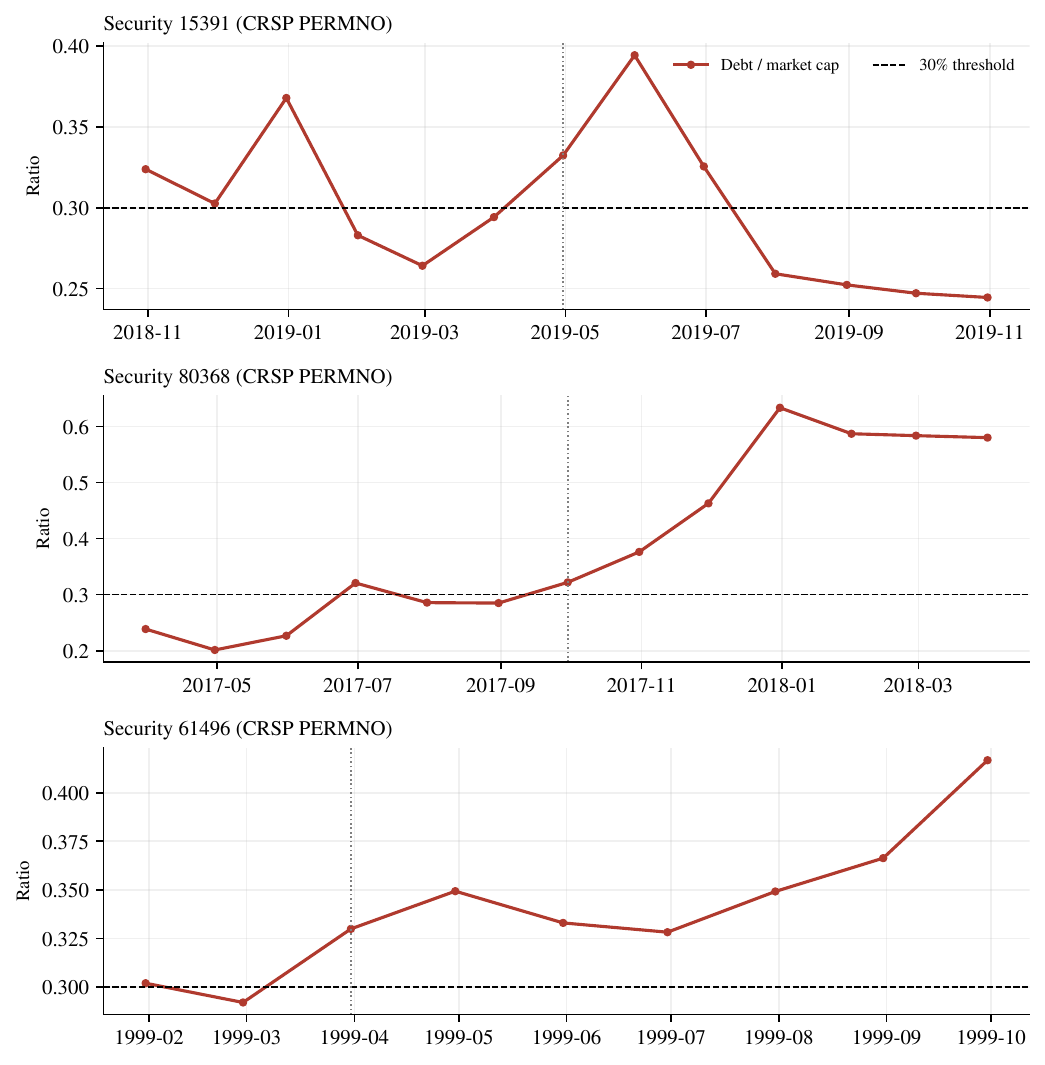}
\caption{Three typical AAOIFI denominator-only debt-screen breaches. Among breaches that move the debt ratio from 25--30\% to 30--40\%, these are the three whose market-capitalisation change is closest to the median. The dotted line marks the first month in the non-compliant state. They illustrate the mechanism and do not establish prevalence.}\label{fig:examples}
\end{figure}

Table~\ref{tab:sustained} requires a reversal to last three consecutive months; Kaplan--Meier estimates of the probability of a first reversal within 12 months, which retain censored events, are 71.9\% and 37.8\% for denominator-only and numerator-only breaches. Figure~\ref{fig:gradient} reports the in-sample breach gradient by CSCI decile. Event-time medians show that continuous states differ before a binary state change and adjust sharply in the first month of the new state (Figure~\ref{fig:event}). Median debt-to-market-capitalisation stays below 30\% around breaches because many AAOIFI breaches are triggered by the cash screen; heterogeneous binding components make aggregate ratio paths difficult to interpret, so the figure remains descriptive and component-specific event files are retained in the replication package. Figure~\ref{fig:examples} illustrates typical denominator-only breaches.

\begin{table}[!htbp]
\centering
\caption{AAOIFI sustained reversal probabilities (percent)}\label{tab:sustained}\begin{tabular}{llrrr}
\toprule
Direction & Driver & 3m & 6m & 12m \\
\midrule
$C\to NC$ & denominator only & 18.2 & 36.5 & 53.1 \\
$C\to NC$ & numerator only & 2.9 & 8.9 & 18.0 \\
$NC\to C$ & denominator only & 16.5 & 35.5 & 53.0 \\
$NC\to C$ & numerator only & 5.6 & 15.9 & 25.4 \\
\bottomrule
\end{tabular}

\tabnotes{A sustained reversal returns to the origin state for three consecutive observed months, completed within the stated horizon. Denominators require complete follow-up through the horizon.}\end{table}

Table~\ref{tab:lpm} reports the full-score linear probability models summarised in Section~\ref{sec:information}. These estimates are partly mechanical, because each target rule's inputs enter CSCI; Table~\ref{tab:incremental} removes the target rulebook and controls for its own margin.

\begin{table}[!htbp]
\centering
\caption{CSCI and next-month breaches}\label{tab:lpm}\begin{tabular}{lrrrr}
\toprule
Standard & $N$ & Effect (pp) & SE (pp) & $t$ \\
\midrule
AAOIFI & 354,469 & -3.39 & 0.14 & -24.61 \\
DJIM & 353,956 & -1.41 & 0.08 & -18.42 \\
S\&P & 372,427 & -1.30 & 0.08 & -16.52 \\
FTSE & 322,406 & -0.27 & 0.06 & -4.32 \\
MSCI main & 300,521 & -0.28 & 0.06 & -4.47 \\
MSCI M-series & 355,481 & -1.35 & 0.07 & -18.64 \\
SC Malaysia & 364,041 & -0.31 & 0.06 & -4.75 \\
\bottomrule
\end{tabular}

\tabnotes{Coefficients are percentage-point changes for one within-sample standard deviation. Each row is a separate linear probability model with firm and month fixed effects and two-way clustered standard errors.}\end{table}

\section{Additional portfolio evidence}\label{sec:port}

\begin{figure}[!htbp]
\centering\includegraphics[width=.76\textwidth]{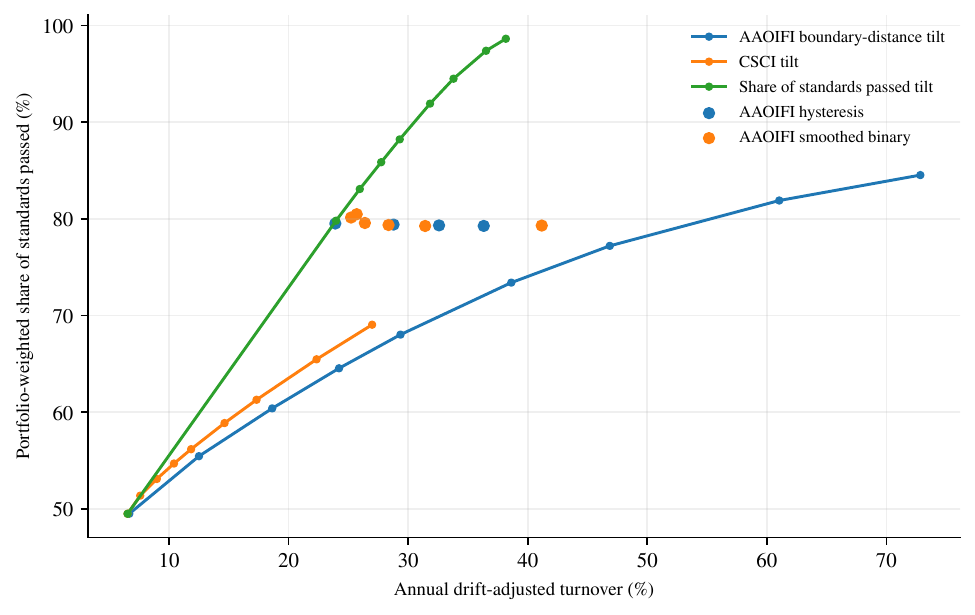}
\caption{Compliance--turnover frontiers. Continuous tilt curves vary the exponent from zero to eight; points show hysteresis and denominator-smoothed binary rules. All signals and weights are lagged one month.}
\label{fig:frontier}
\end{figure}

Figure~\ref{fig:frontier} compares continuous tilts with hysteresis and denominator-smoothed binary rules at matched compliance (Section~\ref{sec:flexible}).

The composite leans toward the four market-capitalisation rulebooks (Section~\ref{sec:disagreement}), so Table~\ref{tab:blocks} rebuilds the tilt portfolios of Table~\ref{tab:portfolios} from the market-capitalisation block alone and from the asset-based block alone. Turnover is similar across scores. The market-capitalisation block reproduces the primary results. The asset-based tilts have lower Sharpe ratios, and the squared asset-based tilt has a negative six-factor alpha of $-0.86\%$ a year ($t=-2.05$). Tilting toward low debt and cash relative to total assets therefore does not deliver the same risk-adjusted performance as tilting toward low debt relative to market value; we report this rather than choose between the scores on performance.

\begin{table}[!htbp]
\centering
\caption{Tilt portfolios built from market-capitalisation and asset-based rulebook blocks}\label{tab:blocks}\begin{tabular}{llrrrrr}
\toprule
Score & $\kappa$ & Turnover & Sharpe & Alpha & $t$(alpha) & Net Sharpe \\
\midrule
All seven rulebooks (primary) & 1 & 9.0 & 0.493 & -0.08 & -0.27 & 0.490 \\
 & 2 & 11.8 & 0.484 & -0.23 & -0.70 & 0.480 \\
Market-cap rulebooks only & 1 & 9.6 & 0.495 & 0.10 & 0.34 & 0.492 \\
 & 2 & 13.1 & 0.488 & 0.11 & 0.33 & 0.484 \\
Asset-based rulebooks only & 1 & 9.1 & 0.462 & -0.48 & -1.45 & 0.459 \\
 & 2 & 11.7 & 0.450 & -0.86 & -2.05 & 0.447 \\
\bottomrule
\end{tabular}

\tabnotes{Weights are proportional to lagged market capitalisation times the lagged score raised to $\kappa$, over permissible-sector securities with an observed score; 311 months. The market-capitalisation block averages the AAOIFI, DJIM, S\&P and MSCI M-series standard scores, and the asset-based block averages FTSE, MSCI main and SC Malaysia. Turnover and alpha are annualised percentages; alpha uses six factors with HAC(6) errors. Net Sharpe deducts 50 basis points per round trip.}\end{table}

\section{Firm, support and accounting-timing robustness}\label{sec:firm}

Selecting one primary security per firm gives 35.87\% disagreement, a 66.07\% denominator-only share of AAOIFI breaches, and a 49.66\% three-month reversal rate; the corresponding firm-level-market-cap values are 36.60\%, 66.08\%, and 49.69\%. Complete-seven support in the latter design is 746,637 observations because firm-level trailing windows are rebuilt from scratch, which accounts for its somewhat higher disagreement. Delaying the entry of each statement by another month or capping its life at 12 rather than 18 months after entry changes none of those quantities by more than 0.6 percentage points. On firm-level market capitalisation, annual turnover remains 8.4\% and 11.3\% for the two CSCI tilts. Pairwise-available disagreement differs from the complete-seven-rulebook estimates by at most 0.24 percentage points. Table~\ref{tab:firmrobust} reports the compact comparison.

\begin{table}[!htbp]
\centering
\caption{Firm, support, and accounting-timing robustness}\label{tab:firmrobust}\fittable{\begin{tabular}{lrrrrrrr}
\toprule
Design & Disagree & Denom. & Reverse & Spot & 24m & $\kappa=1$ & $\kappa=2$ \\
\midrule
Baseline security-month & 35.85 & 66.1 & 49.7 & 2.51 & 1.03 & 9.0 & 11.8 \\
Primary security per firm & 35.87 & 66.1 & 49.7 & 2.51 & 1.03 & 8.9 & 11.8 \\
Firm-level market cap & 36.60 & 66.1 & 49.7 & 2.51 & 1.04 & 8.4 & 11.3 \\
Accounting lag +1 month & 35.92 & 65.5 & 50.2 & 2.50 & 1.03 & 9.0 & 11.9 \\
Accounting expires at 12 months & 35.85 & 66.2 & 49.7 & 2.51 & 1.03 & 9.1 & 12.0 \\
\bottomrule
\end{tabular}
}
\tabnotes{Numeric entries are percentages. Denom. is the denominator-only share of AAOIFI $C\to NC$ changes; Reverse is its three-month return-to-origin probability. Spot and 24m are common-support debt-screen transition frequencies. The last two columns are annual drift-adjusted CSCI-tilt turnover. Firm-level market capitalisation sums contemporaneous securities within firm; primary-security rows retain the largest security each firm-month.}\end{table}

\section{Partial official-list benchmark}\label{sec:sc}
We use the Securities Commission Malaysia's May and November compliant lists, specifically the 25 lists from November 2013, when the SC introduced its quantitative financial-ratio benchmarks, through November 2025. Every list through November 2020 reproduces the number of securities printed in the SC's own document; later documents print no total. Stock codes are linked to Compustat Global through the ordinary-share ISIN, which embeds the Bursa Malaysia code, separately for each list date because codes are reassigned. The benchmark is intentionally restricted: it covers securities whose code appears on at least one list, requires an active Compustat Global security within 45 days before each list date, and attaches the latest annual accounts after a six-month lag with an 18-month stale-data limit. Observable 33\% debt/assets and cash/assets tests plus an SIC sector proxy agree with official list membership in 74.7\% of 18,528 security-list observations; sensitivity is 74.8\%, specificity 74.2\%, and precision 94.7\%. Agreement falls from 83\% for the November 2013 list to 68\% for November 2025. Among officially compliant securities, the share failing the public debt test rises after 2019 and the share failing the cash test rises steadily, which is consistent with the SC's use of conventional rather than total debt and cash and with lease liabilities entering reported debt from 2019; we cannot separate these explanations. These are diagnostics, not full-provider validation. The universe omits securities that never appear on an official compliant list, while official decisions use business-revenue and qualitative inputs absent from Compustat. False positives and false negatives can also arise from Shariah-specific debt and cash adjustments, interim accounts, source-vintage differences, and mapping error. We therefore make no stronger external-validation claim and retain the confusion matrices only in the replication tables.

\section{Fund-holdings checks}\label{sec:funds}

\subsection*{Index ETFs}

We compare emulated breaches with the decisions of two exchange-traded funds that fully replicate Shariah indices, using their quarter-end holdings in SEC Form N-PORT filings from May 2020 to December 2024. SPUS follows S\&P's screens and HLAL FTSE's. The test and its hypotheses were specified before exits were linked to emulated states. An emulated breach of the fund's rule raises the probability that the fund sells the security by the next snapshot: from 1.5\% to 4.9\% for SPUS (Fisher $p=0.066$) and from 3.5\% to 8.7\% for HLAL ($p=0.014$). Yet only 7.7\% and 13.0\% of exits coincide with an emulated breach. Most exits presumably reflect provider data, buffers, index reconstitution, or other inputs that public emulations do not capture, and most emulated breaches of held securities come from the approximate income screen, none of which coincided with a sale. Denominator-driven breaches are almost absent among these funds' large-capitalisation holdings: none for SPUS, whose denominator is a 36-month average of market capitalisation, and three, all driven by total assets, for HLAL. This is consistent with the low denominator sensitivity of averaged and asset-based rules (Table~\ref{tab:drivers}). The architecture and reversal predictions therefore cannot be tested with these funds; an AAOIFI-following fund with disclosed holdings would be required.

The check uses all quarter-end N-PORT holdings of SPUS (SP Funds S\&P 500 Sharia Industry Exclusions ETF) and HLAL (Wahed FTSE USA Shariah ETF) whose report dates fall between May 2020 and December 2024, linked to CRSP through historical CUSIPs. The primary risk set contains securities held at one snapshot that are in the panel and in a permissible sector, are emulated compliant under the fund's rule at that month, and, for SPUS, are S\&P 500 members. Exits are excluded as non-screen events when the security has no panel observation at the next snapshot or, for SPUS, has left the S\&P 500; the one SPUS snapshot pair spanning the September 2023 rule change is excluded. HLAL's parent-index membership is not observed, so its reconstitution exits remain in the sample. An emulated breach is a non-compliant state under the fund's rule at the next snapshot, and its driver applies the decomposition of Section~\ref{sec:drivers} between the two snapshot months. Hypotheses, exclusions and tests were recorded before exits were linked to emulated states; all results are reported. Because no exit follows a denominator-only breach, the pre-specified comparisons of driver composition across funds and of re-entry by driver cannot be estimated. Round trips are exits followed by re-entry within four quarters, valued at the last reported position. SPUS tracks the S\&P 500 Shariah Industry Exclusions Index, which is drawn from the S\&P 500 Shariah Index and governed by S\&P Dow Jones Indices' Shariah methodology; its accounting screens divide by the 36-month average market value of equity throughout our window, with the September 2023 change described in Section~\ref{sec:setting} \citep{SP2024}. A 2019 registration filing for the fund described FTSE-type ratios to total assets \citep{SPFunds2019}, but the effective prospectuses describe the screens only in general terms and defer to the index provider, so we follow the index methodology. The index's additional industry exclusions (aerospace and defence, financial exchanges and data, and payment processing, which replaced data processing in March 2023) are not emulated; exits they cause are non-screen exits that stay in the base rate.

\begin{table}[!htbp]
\centering
\caption{Emulated breaches and actual exits of two Shariah index ETFs}\label{tab:etf}\fittable{\begin{tabular}{lrr}
\toprule
 & SPUS (S\&P rules) & HLAL (FTSE rules) \\
\midrule
\multicolumn{3}{l}{\textit{Panel A. Exits and emulated breaches}} \\
Snapshot pairs & 18 & 18 \\
Held security-quarters, primary risk set & 2,531 & 1,812 \\
Exits in primary risk set & 39 & 69 \\
Exit rate with emulated breach in the interval, \% ($N$) & 4.9 (61) & 8.7 (103) \\
Exit rate without emulated breach in the interval, \% ($N$) & 1.5 (2,470) & 3.5 (1,709) \\
Odds ratio (Fisher $p$) & 3.50 (0.066) & 2.63 (0.014) \\
Share of exits with an emulated breach in the same interval, \% & 7.7 & 13.0 \\
\midrule
\multicolumn{3}{l}{\textit{Panel B. Emulated breaches of held securities by driver: breaches (exits)}} \\
Income screen & 47 (0) & 25 (0) \\
Numerator only & 10 (2) & 60 (7) \\
Denominator only & 0 (0) & 3 (0) \\
Both sufficient & 0 (0) & 3 (1) \\
Joint & 0 (0) & 3 (1) \\
Mixed components & 3 (1) & 8 (0) \\
Sector & 1 (0) & 1 (0) \\
\midrule
\multicolumn{3}{l}{\textit{Panel C. Re-entry within four quarters and round trips}} \\
Re-entry after breach-linked exits, \% ($N$) & 33.3 (3) & 0.0 (9) \\
Re-entry after other exits, \% ($N$) & 19.4 (36) & 11.7 (60) \\
Annual round-trip value, \% of net assets & 0.17 & 0.24 \\
\bottomrule
\end{tabular}
}
\tabnotes{Quarterly N-PORT holdings, May 2020--December 2024. Emulated states use the corrected S\&P and FTSE rules of Table~\ref{tab:rulebooks}. Panel B counts emulated breaches of held securities in the primary risk set, with exits in parentheses; the HLAL denominator is total assets. Fisher tests are two-sided.}\end{table}

\subsection*{Active manager: Saturna Capital (Amana Funds)}

The ETF check cannot test the spot-denominator results, so we add an exploratory check on an active manager. Saturna Capital manages the Amana Funds, the largest U.S. Shariah-screened mutual funds. Their financial screens limit debt to 33\% of a 12-month trailing average of market capitalisation and receivables to 45\% of total assets \citep{Saturna2019,Saturna2021}. We use Saturna's quarterly 13F holdings from December 2004 to December 2024, exported from LSEG Workspace. The design and hypotheses were recorded after the ETF results were known but before any holding was linked to an emulated state (Table~\ref{tab:saturna}). Emulated breaches of the Amana rule again raise the exit probability: from 0.4\% to 2.7\% for positions of at least 0.10\% of the portfolio (Fisher $p=0.034$, from only 14 exits) and from 3.2\% to 8.8\% across all positions ($p<0.001$). Only 14.3\% and 7.5\% of exits coincide with a breach. Because breach and exit are observed over the same quarter, these are breach-associated exits; quarterly holdings cannot show which came first. Breaches already visible in the first month of a quarter come with somewhat higher exit rates than breaches that appear later (13.3\% against 7.8\% across all positions), but the difference is not significant. We predicted that AAOIFI breaches which the fund's smoothed rule does not register would not coincide with sales. In raw rates this prediction fails: among positions that remain Amana-compliant, an AAOIFI breach raises the exit rate from 2.8\% to 11.7\% across all positions. These breaches coincide with large price falls, however. The median AAOIFI denominator-only breach comes with a quarterly return of about $-20\%$, and managers sell losing positions for reasons unrelated to screens. In an exploratory regression with return controls, the AAOIFI-only effect falls to 2.0 percentage points (standard error 2.8), while the Amana-breach effect remains 4.7 (1.8). We also predicted that denominator-only breaches would coincide with fewer sales. The data point the other way: 9 of 31 Amana breaches driven by the 12-month market-cap denominator end in an exit, against 2 of 40 numerator-only breaches. These are breaches of an averaged denominator, which reflect persistent price declines rather than the transient spot movements that reverse, and the comparison rests on few events. Two further checks address the 13F proxy. The Amana Growth and Amana Income funds' own N-PORT filings, available from November 2019, show that 97.5\% of the value in Saturna's 13F holdings sits in securities those two funds hold. The fund-level filings, however, contain too few breaches to test the hypotheses: across 40 fund-quarter pairs to November 2024 there are 21 Amana breaches, 17 of them from the income proxy, with one exit, and 12 AAOIFI breaches that the Amana rule does not register, with two. Taken together, the checks show that emulated rules capture part of real screening decisions, that most sales have other causes, and that a spot-rule breach adds little information about the sales of a fund that applies a smoothed rule.

Holdings are Saturna Capital Corporation's quarterly positions from the LSEG Workspace Investor Portfolio Report, with current CUSIPs, exported in October 2026. The test uses 13F records from December 2004; earlier records come from semi-annual fund filings. The export's date field is the report period, the quarter end to which a 13F refers, not the SEC submission date: every value is a month end and 99\% are quarter ends. A security is held in a quarter if a record reports that quarter with a positive share position, because the export carries exited and zero-share positions forward with their last report date. The 13F combines the Amana Funds with Saturna's non-Shariah funds and separate accounts; Amana accounts for more than 90\% of reported equity assets in recent filings. The primary risk set therefore requires a portfolio weight of at least 0.10\% at the first quarter of a pair, and all positions form a secondary risk set. Holdings are linked to CRSP by historical and header CUSIPs (mean link rate 91.7\%). The Amana rule combines the permissible-sector screen, debt below 33\% of 12-month trailing average market capitalisation, receivables below 45\% of total assets, and the income proxy below 5\%. Saturna's documents give no revision dates, so the rule is held fixed. Other definitions follow the ETF check. The AAOIFI comparison uses positions compliant under both rules at the first quarter that remain Amana-compliant at the second. The exploratory regressions, added after the pre-specified tests were seen, include quarter fixed effects, return-decile indicators and the negative part of the quarter's compounded CRSP return, with standard errors clustered by firm and quarter. Two quarters in which the number of reported positions falls by more than 25\% (December 2008 and March 2016) are excluded in a robustness row. Panel D splits Amana breaches by whether the security is already non-compliant in the first month after the earlier quarter end. For the fund-level check, N-PORT holdings of Amana Growth and Amana Income are taken from each fund's quarterly filings (fiscal quarters ending February, May, August and November), using the latest filing for each report date, and processed with the same code. The 13F overlap compares each N-PORT date with the next calendar quarter end. The plan, code and outputs are in the replication package; the licensed LSEG export is not redistributed.

\begin{table}[!htbp]
\centering
\caption{Emulated breaches and actual exits of an active Shariah manager}\label{tab:saturna}\fittable{\begin{tabular}{lrr}
\toprule
 & Weight $\geq$ 0.10\% (primary) & All positions \\
\midrule
\multicolumn{3}{l}{\textit{Panel A. Tests fixed before linking}} \\
Quarter pairs, 2005--2024 & 80 & 80 \\
Held security-quarters in risk set & 3,476 & 5,990 \\
Exits in risk set & 14 & 200 \\
Exit rate with Amana breach in the interval, \% ($N$) & 2.7 (74) & 8.8 (171) \\
Exit rate without Amana breach in the interval, \% ($N$) & 0.4 (3,402) & 3.2 (5,819) \\
Odds ratio (Fisher $p$) & 7.85 (0.034) & 2.93 ($<$0.001) \\
Share of exits with an Amana breach in the same interval, \% & 14.3 & 7.5 \\
AAOIFI breach, Amana compliant: exit rate, \% ($N$) & 1.6 (64) & 11.7 (137) \\
No breach under either rule: exit rate, \% ($N$) & 0.3 (3,109) & 2.8 (5,228) \\
Odds ratio (Fisher $p$) & 4.92 (0.201) & 4.54 ($<$0.001) \\
\midrule
\multicolumn{3}{l}{\textit{Panel B. Breaches of held securities by driver: breaches (exits)}} \\
Amana: income screen & 42 (0) & 87 (4) \\
Amana: numerator only & 17 (0) & 40 (2) \\
Amana: denominator only (12-month market cap) & 10 (2) & 31 (9) \\
Amana: both, joint, mixed or sector & 5 (0) & 13 (0) \\
AAOIFI: denominator only (spot market cap) & 41 (1) & 90 (15) \\
AAOIFI: numerator only & 22 (0) & 53 (3) \\
\midrule
\multicolumn{3}{l}{\textit{Panel C. Exploratory: exit LPM with return controls, pp (SE)}} \\
Amana breach & 2.5 (2.0) & 4.7 (1.8) \\
AAOIFI breach, Amana compliant & 1.2 (1.4) & 2.0 (2.8) \\
Amana denominator-only breach & 19.8 (11.4) & 21.8 (7.3) \\
Other Amana breach & $-$0.4 (0.2) & 0.8 (1.5) \\
Amana breach, excluding reporting-drop quarters & -- & 3.7 (1.7) \\
\midrule
\multicolumn{3}{l}{\textit{Panel D. Timing: exit rate by when the Amana breach first appears, \% ($N$)}} \\
Breach visible in the first month after the earlier snapshot & 8.3 (12) & 13.3 (30) \\
Breach appears later in the interval & 1.6 (62) & 7.8 (141) \\
\bottomrule
\end{tabular}
}
\tabnotes{Saturna Capital 13F holdings, quarterly, December 2004--December 2024. Amana breach: non-compliant at the second quarter of a pair under the emulated Amana rule after a compliant first quarter. Panel B counts breaches of held securities in the risk set, with exits in parentheses; AAOIFI rows count all AAOIFI breaches in the risk set. Panel C reports linear probability models of exit, in percentage points; they are exploratory. Fisher tests are two-sided.}\end{table}

\section{Scoped daily denominator diagnostic}\label{sec:daily}
A pre-existing daily cache covering 2018--2024 is used only to ask how much a market-capitalisation ratio can move between accounting updates. Each trading day uses the preceding month-end accounting numerator, and consecutive-day decompositions require an unchanged information month. After the safe join, 4,584,272 daily rows form 221,693 firm-months. Of these, 4.02\% cross the 30\% debt/market-cap boundary intramonth and 0.84\% start and end compliant despite crossing in between. Conditional on an observed daily crossing while debt is fixed, a price-only counterfactual is sufficient for 99.1\% of $C\to NC$ and 96.6\% of $NC\to C$ events. Those conditional percentages are close to mechanical by construction and are not population prevalence estimates. The cache contains firms selected for a separate event study, so this exercise is a timing and implementation diagnostic rather than a headline finding.

\end{document}